\documentclass[jpcm]{iopart}

\usepackage{epsf,citesort,graphicx}

\begin{document}

\title{Competition between glass transition and liquid-gas separation in attracting
colloids}

\author{A. M. Puertas$^1$, M. Fuchs$^2$, M.E. Cates$^3$} 

\address{$^1$ Group of Complex Fluids Physics, Department of Applied Physics, University of Almeria, 04120 Almeria, Spain}
\address{$^2$ Fachbereich Physik, University of Konstanz, D-78457 Konstanz, Germany}
\address{$^3$ SUPA, School of Physics, The University of Edinburgh, JCMB Kings Buildings, Mayfield Road, Edinburgh EH9 3JZ, UK}

\begin{abstract}
We present simulation results addressing the phenomena of colloidal gelation induced by attractive interactions. The liquid-gas transition is prevented by the glass arrest at high enough attraction strength, resulting in a colloidal gel. The dynamics of the system is controlled by the glass, with little effect of the liquid-gas transition. When the system separates in a liquid and vapor phases, even if the denser phase enters the non-ergodic region, the vapor phase enables the structural relaxation of the system as a whole.
\end{abstract}


\section{Introduction}

In colloids with short range attractions, the system ``gels'' at high enough attraction strength. This colloidal gel is a solid, with a low colloid density, stabilized by the bonds between particles. Usually, gelation occurs above crystallization and liquid-gas transition, i.e. increasing the attraction strength, the system enters the fluid-crystal coexistence region, at higher strengths it separates into a liquid and a gas, and finally it gels \cite{poon95,manley05,sedgwick05}. In some cases, gels are found outside the liquid-gas separation boundary \cite{shah03}, although the latter cannot be identified and a clear independence between gelation and phase separation cannot be claimed. Thus, it has been argued that gels are indeed states with arrested phase separation \cite{manley05,foffi05}. The mechanism for arrest can be either the dense phase crossing a glass transition (either attraction or repulsion driven) \cite{foffi05,sastry00}, or a glass transition driven by the particle bonding that prevents the phase separation.

In this work we have simulated a system that mimics the mixture of a colloid with a non-adsorbing polymer (which induces attraction between the colloids), in three states beyond the liquid-gas transition. The two states with stronger attractions indeed show arrested phase separation, and the system has formed a percolating network of particles with voids and tunnels. The dynamics of the system is studied in the three states, and we find that the system with arrested phase separation show many properties typical of glass aging. The system with the lowest attraction strength separates into a dense and a dilute phase, but the liquid phase is non-ergodic, i.e. it has reached the glass region at higher density. The dynamics of the system with a long range repulsive barrier, which suppresses the liquid-gas separation, has been studied previously \cite{puertas05,puertas06}, what allows us to conclude here that gels are indeed caused by a quench to the attractive glass region. Quenches to lower attraction strength result in phase separation -- the liquid may undergo a glass transition and become non-ergodic, but the system does not end up with the typical gel properties.  

\section{Simulation details}

Newtonian dynamics simulations were run for a system comprised by 1000 quasi hard particles with an attractive interaction mimicking the effective interaction between colloids in colloid-polymer mixtures due to polymer depletion. The system is polydisperse with radii distributed according to a flat distribution of width 10\% of the average radius, $a$. The core-core repulsion is given by $V(r) = k_BT (r/\sigma)^{-36}$, where $k_BT$ is the thermal energy and $\sigma=a_1+a_2$. The attractive interaction is given by the simple depletion model developed by Asakura and Oosawa \cite{likos01}, corrected to consider a polydisperse system \cite{mendez00}; the attraction strength is given by the polymer volume fraction, $\phi_p$, and the range by the size of the polymers $2\xi$. Further details of the total interaction potential can be found in previous works \cite{puertas05}. The colloidal density is reported as volume fraction, $\phi_c$, and the attraction strength in units of $\phi_p$; dimensionless units are used by setting the mean radius $a=1$, the thermal velocity $v=\sqrt{4/3}$ and the mass $m=1$. 

For the attraction range and volume fraction used in this work, $2\xi=0.2a$ and $\phi_c=0.40$, respectively, the system undergoes liquid-gas separation at $\phi_p\approx 0.3$ (the crystallization transition occurs in monodisperse systems at lower $\phi_p$). The system is equilibrated without attraction and instantaneously quenched to three different states: $\phi_p=0.35$, $\phi_p=0.50$ and $\phi_p=0.80$. Previous works where the liquid-gas transition is inhibited by means of a long range repulsive barrier showed that there is an attractive glass transition at $\phi_p\approx 0.43$ \cite{puertas05}. Thus, the first state is below the glass point, whereas the other two are above. The structure and dynamics are studied below as a function of the time elapsed since the quench, termed {\sl waiting time}.

\begin{figure}
\begin{center}
\includegraphics[width=0.6\textwidth]{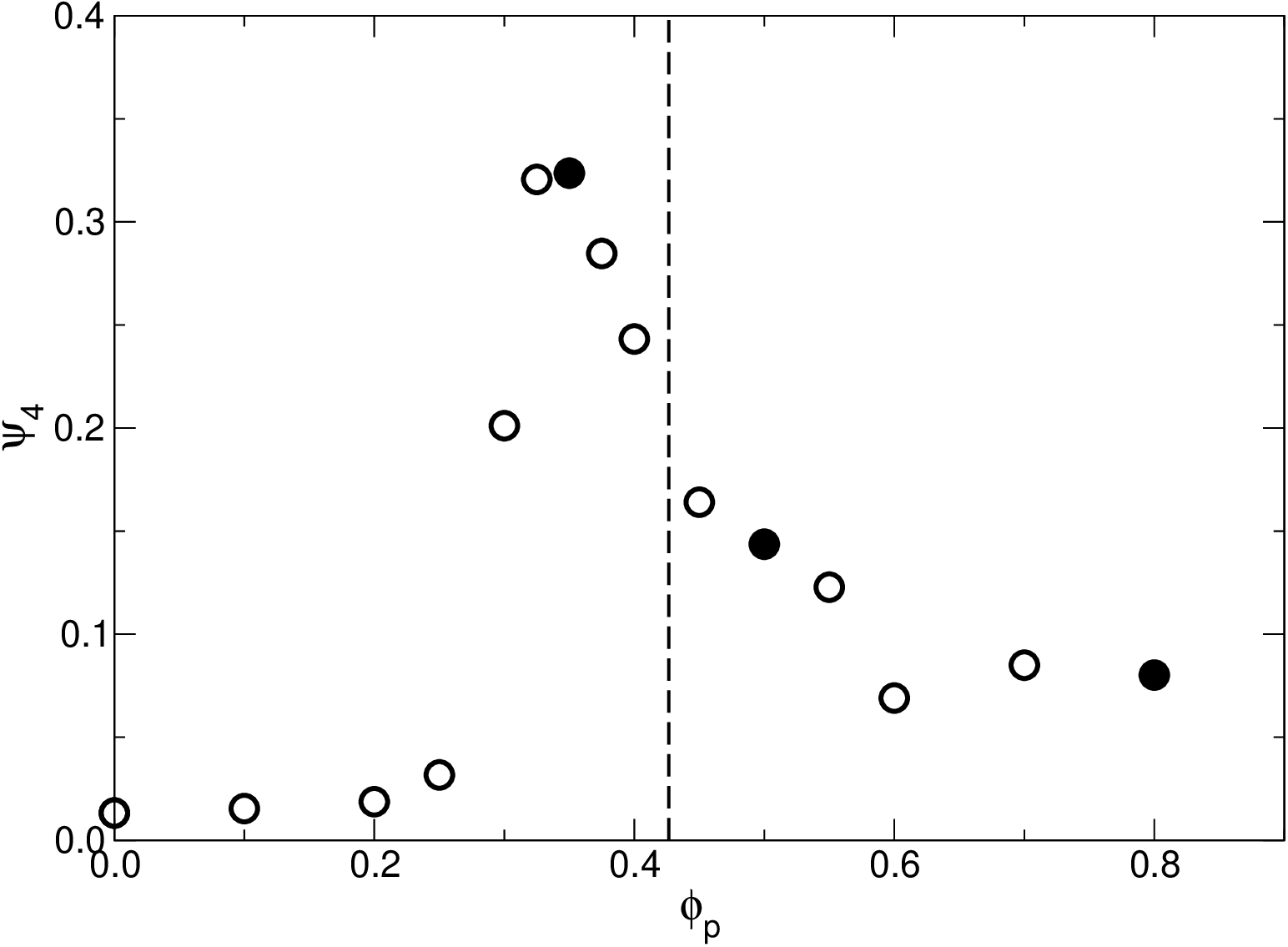}
\end{center}
\caption{\label{psi4} Phase separation order parameter, $\psi_4$, for different states along the isochore $\phi_c=0.40$. The system is divided in $4^3$ boxes, and $\psi_4=\sum_i (\rho_i-\rho)^2$, where $\rho_i$ is the density in every box, and the summation runs over all the boxes. The black circles mark the states studied in detail, $\phi_p=0.35$, $\phi_p=0.50$ and $\phi_p=0.80$. The vertical dashed line shows the state where the glass transition was found without liquid-gas separation.}
\end{figure}

\begin{figure}
\begin{center}
\includegraphics[width=0.6\textwidth]{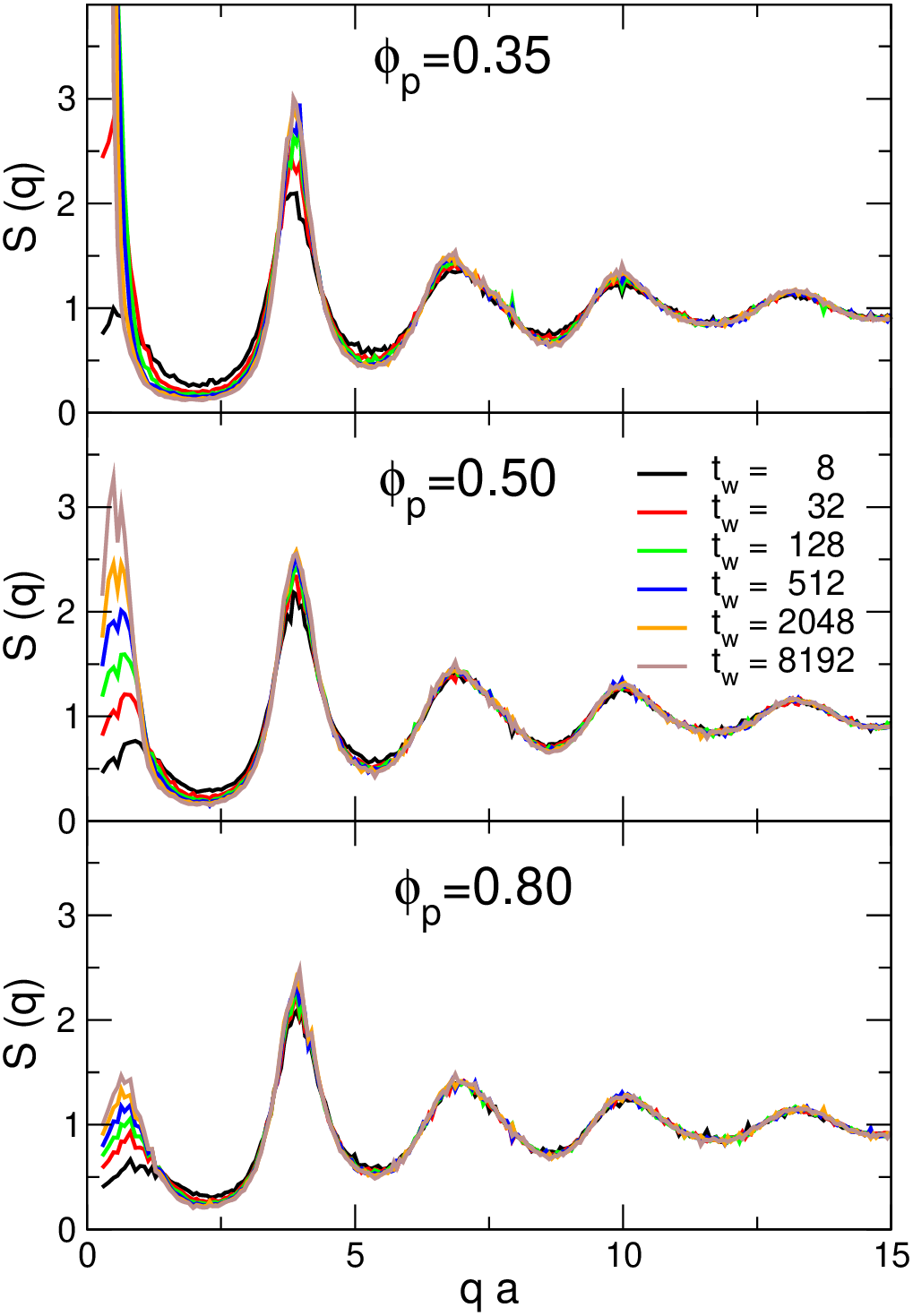}
\end{center}
\caption{\label{sq} Structure factor for the three states as labeled and at the waiting times shown.}
\end{figure}

\section{Results}

Fig. \ref{psi4} shows the evolution of the density inhomogeneity in the system as a function of the attraction strength, $\phi_p$, along $\phi_c=0.40$. Homogeneous fluids are obtained at low $\phi_p$, and liquid-gas separation at $\phi_p \approx 0.30$. However, instead of observing increasing inhomogeneity of the system with $\phi_p$, due to denser liquids and more dilute vapors, we note that the phase separation is impeded by an additional mechanism. We show here that this mechanism is the glass transition already studied in the same system without phase separation (vertical dashed line in the figure).

The inhibition of liquid-vapor separation is also noticeable in the evolution of the structure factor, shown in Fig. \ref{sq}. Whereas the state $\phi_p=0.35$ shows the typical behaviour of spinodal decomposition, with a peak at low wavevectors that grows and moves to lower $q$, the states at higher attraction strength show a more homogeneous structure. Only a peak at low $q$ is observed, which grows much slower than at $\phi_p=0.35$, and which is smaller the higher $\phi_p$. A similar peak at low $q$ is obtained in the system without phase transition, due to local compaction of the system \cite{puertas05}. The arrested phase separation scenario is fully in agreement with Fig. \ref{psi4} and with previous findings in simulations and experiments; the resulting gels are heterogeneous locally and show arrested dynamics \cite{foffi05,manley05,sedgwick05}.
\begin{figure}
\begin{flushright}
\includegraphics[width=0.85\textwidth]{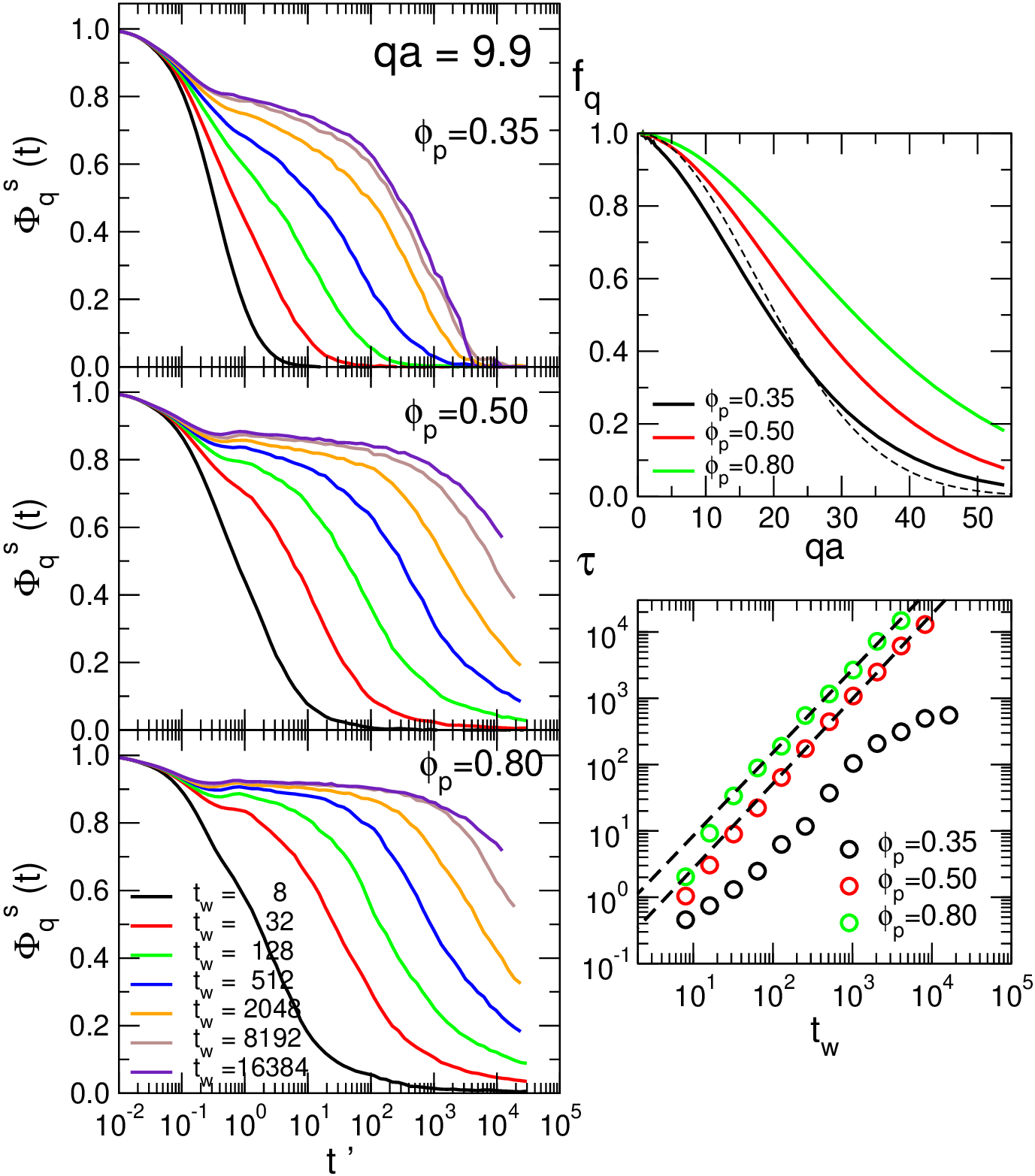}
\end{flushright}
\caption{\label{fsqt} Incoherent density correlation functions for the three states and at different waiting times, as labeled. The right hand panels show the incoherent non-ergodicity parameter, $f_q$, as a function of $q$ (the dashed line is the Gaussian approximation with a localization length equal to the attraction range) and the structural relaxation time, $\tau$ as a function of the waiting time ($\tau$ is defined as $\Phi_q^s(\tau)=f_q/2$, for $qa=9.9$).}
\end{figure}

The dynamics of the system is studied in Fig. \ref{fsqt} by means of the incoherent part of the density correlation function, $\Phi_q^s(t')\:=\: 1/N \sum \exp \left\{ {\bf q} ({\bf r}_i(t)-{\bf r}_i(t_w)\right\}$, where the summation runs over the $N$ particles in the system and $t'=t-t_w$. The density correlation functions in the left panels show the typical features of a glass transition, i.e. a two step decay, where the second, structural, relaxation increases with waiting time. The height of the intermediate plateau, the non-ergodicity parameter, is presented in the right upper-most panel as a function of $q$, and the evolution of the relaxation time with waiting time is given in the lower panel. 

The state with $\phi_p=0.35$ shows the fastest relaxations and the lowest non-ergodicity parameters (yielding a localization length longer than the attraction range), but more importantly, the time scale for structural relaxation saturates. On the other hand, the dynamics of the states for higher $\phi_p$ do not saturate and $\tau$ follows a power law with waiting time, with exponents larger than one, and the localization length is smaller than the interaction range. Similar behaviour was observed in aging of the attractive glass obtained in this system without phase separation \cite{puertas06}. 
\begin{figure}
\begin{flushright}
\includegraphics[width=0.85\textwidth]{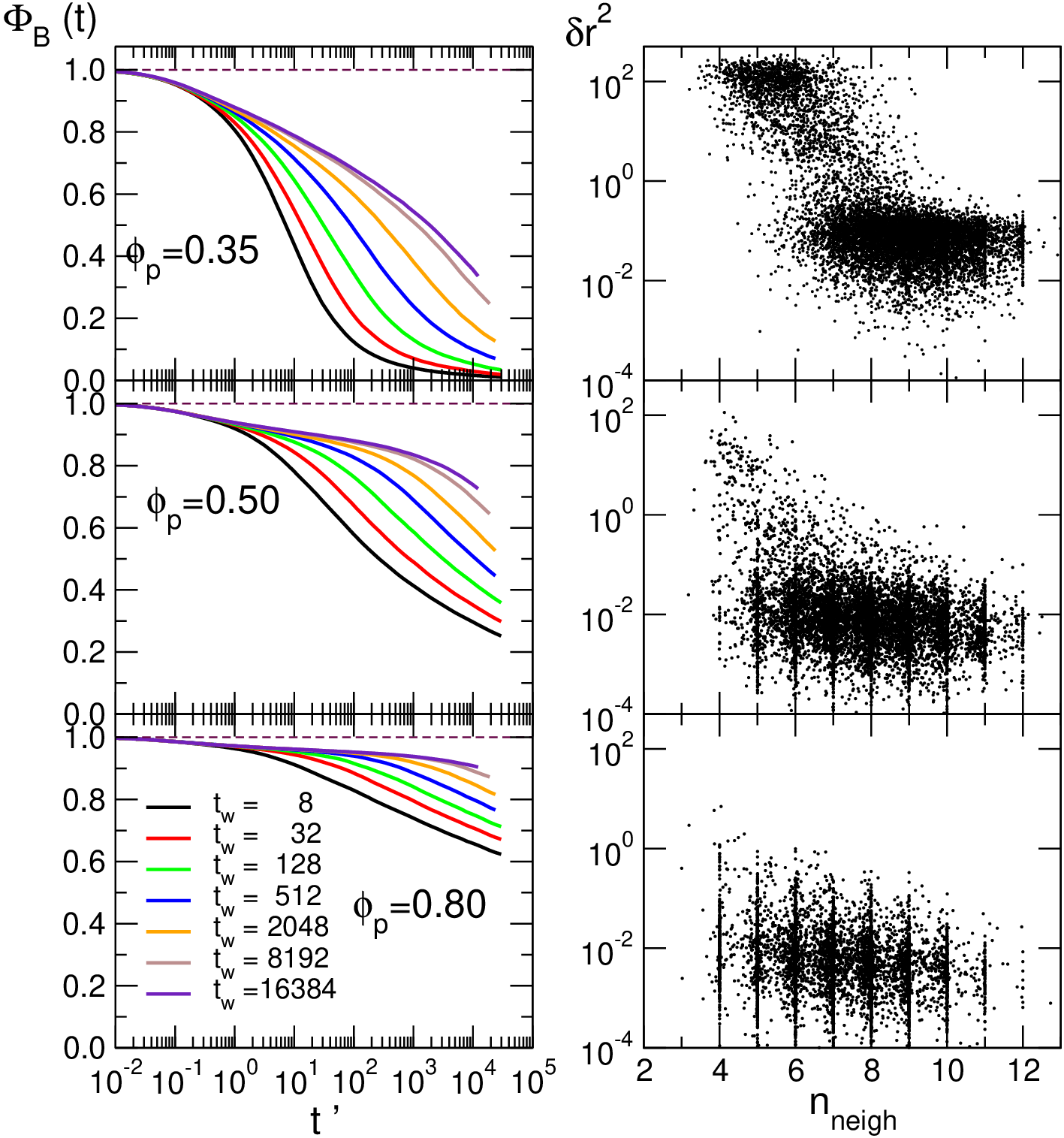}
\end{flushright}
\caption{\label{bonds} Left panels: Bond correlation functions for the three states and waiting times labeled. Right panels: Correlation between the squared displacement between $t'=0$ and $t'=1000$ and the average number of neighbours in this time interval for $t_w=16384$. }
\end{figure}

The dynamical arrest is caused by the bonds between particles due to the attraction, as observed by the non-ergodicity parameters. It is thus interesting to study the bond correlation function, i.e. the fraction of bonds that have existed uninterruptedly since $t_w$ until $t=t_w+t'$, Fig. \ref{bonds} (two particles are bonded when their separation is smaller than the attraction range). In agreement with the dynamics studied in Fig. \ref{fsqt}, the bonds are stronger and live longer for high $\phi_p$. Note that even at the highest $\phi_p$ studied, $\phi_p=0.80$, the bonds are still reversible, and $\Phi_B(t')$ does not show any plateau. The correlation between the displacement of a given particle and its mean number of bonds (neighbours), shown in the right-hand panels, indicates that particles with a small number of neighbours (in average) move much longer than average, but this population of mobile particles is only apparent in the state $\phi_p=0.35$. These particles, thus, comprise the vapor phase which is in coexistence with the liquid phase, with much lower mobility.

The overall relaxation of the system observed in the density correlation function for $\phi_p=0.35$ is therefore caused by the vapour phase, due to the exchange between particles in both phases, although the liquid one can be itself inside the non-ergodic region (most probably in the attractive glass, according to the non-ergodicity parameter). A population of {\sl fast} particles was also observed in the system without phase separation and aided the structural relaxation below the glass transition. 
The states $\phi_p=0.50$ and $\phi_p=0.80$, on the other hand, are quenched above the glass transition (found in the system without phase separation \cite{puertas05}) and the liquid-gas transition is inhibited. Here, no vapor phase is present, and the system cannot relax structurally, also in agreement with previous findings in the system without phase separation (there, the population of fast particles vanished in the glass states \cite{puertas06b}). Thus, the dynamics of the system is controlled by the glass transition, with only small effects due to the phase separation.

Our results show, therefore, that systems undergoing phase separation, where the liquid phase enters the non-ergodic region, may still relax and appear ergodic due to the vapor phase. In order to observe really arrested dynamics and phase separation, the system must be quenched beyond the glass transition, to prevent the formation of the gas phase. We cannot say, however, if the crossover from the ``apparent ergodic'' regime to the  ``arrested'' one, is abrupt or continuous, although our results of static properties point to the latter (Fig. \ref{psi4}). In addition, the glass transition is found with the aid of local compaction of the system, both in the system with and without phase transition, which is also found in other studies of the attractive glass \cite{manley05,zaccarelli06}. Our results, however, cannot elucidate whether this compaction is a real necessity for the glass transition or only favors it.

\ack

Financial support is acknowledged from the M.E.C. -- projects MAT2003-03051-CO3-01 and HA2004-0022 (A.M.P.). We thank F. Sciortino and E. Zaccarelli for useful discussions.

\end{document}